\documentclass[%
 aip,
 amsmath,amssymb,
 reprint,%
]{revtex4-1}

\usepackage{graphicx}
\usepackage{dcolumn}
\usepackage{bm}

\usepackage{hyperref}
\hypersetup{colorlinks=true, urlcolor= blue, citecolor=blue, linkcolor= blue, bookmarks=true, bookmarksopen=false}

\usepackage{amsmath,amssymb}
\usepackage{enumitem}

\usepackage[utf8]{inputenc}
\usepackage[T1]{fontenc}
\usepackage{mathptmx}
\usepackage{etoolbox}

\makeatletter
\def\@email#1#2{%
 \endgroup
 \patchcmd{\titleblock@produce}
  {\frontmatter@RRAPformat}
  {\frontmatter@RRAPformat{\produce@RRAP{*#1\href{mailto:#2}{#2}}}\frontmatter@RRAPformat}
  {}{}
}%
\makeatother
\begin{document}

\preprint{AIP/123-QED}

\title[]{CoF$_3$: a g-wave Altermagnet}
\author{Meysam Bagheri Tagani}
 \affiliation{Department of Physics, University of Guilan, P. O. Box 41335-1914, Rasht, Iran}

 \email{m{\_}bagheri@guilan.ac.ir}

\date{\today}

\begin{abstract}
Altermagnetism, a novel magnetic phase bridging ferromagnetism and antiferromagnetism, exhibits zero net magnetization due to its unique alternating spin arrangements, which cancel out macroscopic magnetization. This phase is characterized by robust time-reversal symmetry breaking and spin-momentum locking, leading to distinct electronic properties advantageous for spintronic applications. In this study, we explore the possibility of altermagnetism in cobalt trifluoride (CoF$_3$) using density functional theory (DFT) with Hubbard U correction combined with spin group theory. Our findings reveal that CoF$_3$ exhibits zero net magnetization similar to a g-type antiferromagnet but with spin degeneracy breaking without spin-orbit coupling, akin to a ferromagnet. The optimized structure of CoF$_3$, characterized by a rhombohedral lattice with centrosymmetric symmetry group R3c, shows significant spin splitting in both valence and conduction bands, reaching up to 45 meV. This spin splitting is attributed to the electric crystal potential and the anisotropy of the spin density, leading to the breaking of Kramers degeneracy.
\end{abstract}

\maketitle

	Altermagnetism represents a novel magnetic phase that bridges the gap between ferromagnetism and antiferromagnetism \cite{vsmejkal2022beyond, vsmejkal2022emerging}. Unlike conventional magnets, altermagnets exhibit zero net magnetization due to their unique spin arrangements, which alternate in a manner that cancels out macroscopic magnetization. This phase is characterized by robust time-reversal symmetry breaking and spin-momentum locking, leading to distinct electronic properties that are advantageous for spintronic applications \cite{gonzalez2023spontaneous, guo2024direct}. The discovery of altermagnetism has opened new avenues for research, particularly in understanding how these materials can be utilized in next-generation magnetic memory devices and other advanced technologies.
	
	Recent experimental studies have provided compelling evidence for the existence of altermagnetism. For instance, thin films of manganese telluride (MnTe) have demonstrated zero net magnetization and a spin-split band structure, confirming the theoretical predictions of altermagnetic behavior \cite{lee2024broken}. Feng et al. reported anomalous Hall effect in an altermagnet RuO$_2$ \cite{feng2022anomalous}. Spin splitter torque was observed in collinear antiferromagnetic RuO$_2$ attributed to the altermagnetism \cite{karube2022observation}. Reimers et al. reported the observation of altermagnetic band splitting in CrSb thin films \cite{reimers2024direct}.  Krempaský and co-workers reported the altermagnetic lifting of Kramers degeneracy in MnTe crystals \cite{krempasky2024altermagnetic}. Theoretical models have further elucidated the underlying mechanisms of altermagnetism, highlighting the role of specific crystallographic symmetries and spin-orbit coupling in stabilizing this unique magnetic phase \cite{vsmejkal2022beyond, vsmejkal2022emerging, das2024realizing, jaeschke2024supercell}.
	
	The unique properties of altermagnets make them promising candidates for various technological applications. In spintronics, altermagnets could enable more efficient spin-based devices due to their strong spin polarization and minimal crosstalk between bits \cite{chen2024emerging}. Additionally, the ability to control altermagnetic order at the nanoscale opens up possibilities for high-density data storage and advanced quantum computing systems \cite{ amin2024altermagnetism}. Future research will likely focus on discovering new altermagnetic materials, optimizing their properties for practical use, and exploring their potential in other fields such as thermoelectricity and quantum sensing \cite{bai2023efficient}, and piezoelectric altermagnetism \cite{ guo2023piezoelectric}.

	Lee et al. synthesized the cobalt triflouride, CoF$_3$, experimentally, and investigated its magnetic properties with high resolution neutron powder diffraction \cite{lee2018weak}.  They found that unlike predicted g-type antiferromagnetic structure, the CoF$_3$ displays an additional magnetic anomaly. They attributed it to the possible weak ferromagnetism induced by spin tilting and Dzyalonski-Moriya interaction. Motivated with this experimental observation, we explore the possibility of altermagnetism in CoF$_3$. We employed the density functional theory (DFT) with Hubbard U correction combined with spin group theory to show that the magnetism observed in the experiment can be attributed to the time reverse breaking induced with spin band splitting. Indeed, CoF$_3$ crystal has zero net magnetization like a g-type antiferromagnet but with spin degeneracy breaking without spin-orbit coupling like a ferromagnet. 
\begin{figure}
\includegraphics[width=0.45\textwidth]{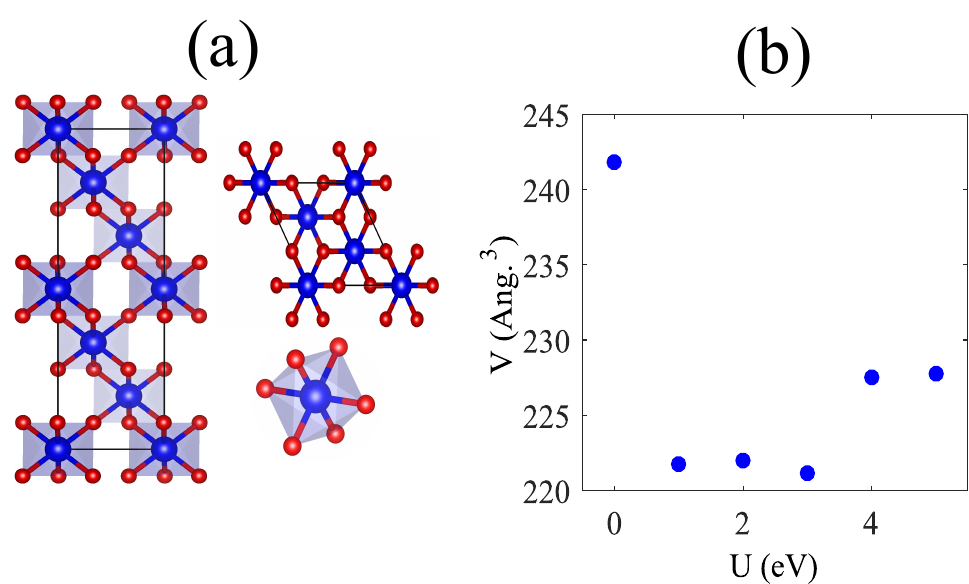}
\caption{\label{Fig1} (a) Views of the CoF$_3$ crystal along the [001] and [110] directions, with the Co$_1$F$_6$ octahedron also depicted. (b) Dependence of the unit cell volume on the Hubbard term.}
\end{figure}

All density functional theory (DFT) calculations were performed using the Quantum Espresso computational code \cite{giannozzi2017advanced}. The k-space was meshed with a grid of 2$\pi$$\times$0.04 \AA$^{-1}$. The electron-ion interaction was described using projected-wave augment pseudopotentials (PAW), and the exchange-correlation interaction was considered using the PBE functional within the GGA approximation \cite{perdew1996generalized}. The convergence criteria for forces and energy in the calculations were set to 0.01 eV/\AA, and 10$^{-7}$ eV, respectively. Due to the strong electron correlation in cobalt atoms, the Hubbard correction within the LDA+U approach was used \cite{cococcioni2005linear}. We considered various U values of 1, 2, 3, 4, and 5 eV to describe the localization of the d orbitals of cobalt atoms. The system was optimized for each U value.

Figure \ref{Fig1}a illustrates the optimized structure of CoF$_3$, characterized by a rhombohedral lattice with a centrosymmetric symmetry group $R3c$. The lattice constants are determined to be a=4.57\AA~ and c=12.56\AA~ for U=5eV, which align well with experimental results obtained from neutron powder diffraction \cite{lee2018weak}. Each cobalt atom is coordinated by six fluorine atoms, forming a perfect octahedral arrangement Co$_1$F$_6$ (see Fig. \ref{Fig1}a). The findings suggest that the lattice constant and unit cell volume exhibit a weak dependence on the Hubbard term. When the correlation of d-orbital electrons is disregarded, the unit cell volume reaches a maximum and stabilizes at a constant value for U=4 and 5 eV. The F-Co-F bond angles show minimal dependence on U, maintaining values of 180 and 90 degrees. The symmetry group of the structure remains invariant with respect to the value of U. Consequently, the primary results for U=5eV are presented in the main text, while results for other U values are provided in the Supplementary Information file.

\begin{figure}
	\includegraphics[width=0.45\textwidth]{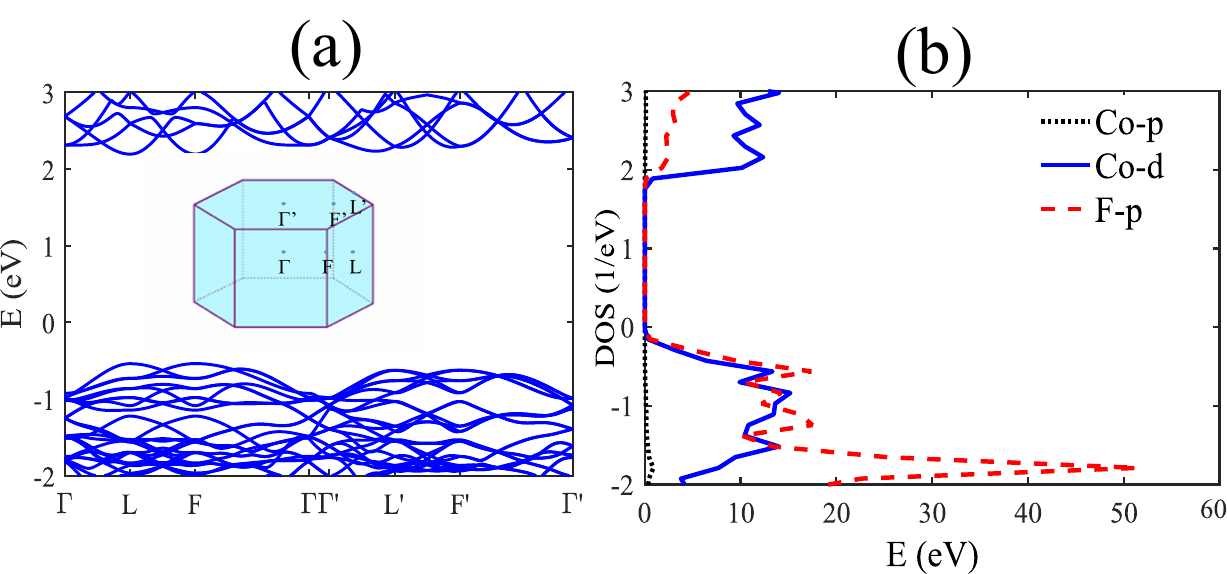}
	\caption{\label{Fig2} (a) Band structure of non-magnetic CoF$_3$, with the inset displaying the first Brillouin zone and its high symmetry points. (b) Partial density of states of the sample.}
\end{figure}

The non-magnetic band structure of CoF$3$ and the partial density of states (PDOS) are examined in Fig. \ref{Fig2}. Additionally, the contributions of various d orbitals of the cobalt atom are plotted in Fig. S1. The non-magnetic band gap of the material is found to be 2.7 eV, considering U=5 eV. The valence band maximum (VBM) and conduction band minimum (CBM) are located at the L point (0.5, 0, 0). Examination of the density of states reveals that the VBM consists of a hybrid of the p$z$ orbitals of fluorine atoms and the d${z^2}$ orbitals of cobalt atoms. In contrast, the CBM is exclusively associated with the cobalt atoms. The hybridization of the d${xz}$ and d$_{yz}$ orbitals is responsible for forming the conduction band edge. The contribution of the d orbitals around the Fermi level indicates that if the sample is antiferromagnetic (AM), we will be dealing with a g-wave.

Nonrelativistic collinear magnets can be described by three different spin groups. Spin groups are expressed by a direct product of spin-only group, $\mathbf{r}_s$, and nontrivial spin groups, $\mathbf{R}_s$ \cite{litvin1974spin}.  Altermagnet, the third phase of collinear magnets, has a net zero magnetic momentum like an antiferromagnet but non-degenerated spin polarized band structure in some part of the Brillouin zone breaking time reversal symmetry like a ferromagnet. The nontrivial spin group describing this class of the magnetic phase is given by \cite{vsmejkal2022beyond}: 
\begin{equation}
\mathbf{R}_s^{III}=[E||\mathbf{H}]+[C_2||A][E||\mathbf{H}]=[E||\mathbf{H}]+[C_2||\mathbf{G-H}],
\end{equation}
where $\mathbf{G}$ are the crystallographic Laue groups. $\mathbf{H}$ is a halving subgroup of G containing only real space transformations between same spin sublattices. In contrast, coset $\mathbf{G-H}=A\mathbf{H}$ is created by transformations $A$ that are only proper or improper rotations contains only real space transformations between opposite spin sublattices. The AM phase includes crystals in which the opposite spin sublattices are connected by rotation not connected by translation or inversion \cite{vsmejkal2022beyond, vsmejkal2022emerging}. 

  CoF$_3$ crystal has $3\overline{m}$ crystallographic Laue group, and its halving subgroup is $\overline{3}$ that interchanges the same spin sublattices. The generator $A$ that interchanges the opposite spin sublattices for the CoF$_3$ is C$21$. Therefore, the nontrivial spin group of the CoF$_3$ crystal is $\overline{3}^2m$  with 12 symmetry elements. Therefore, CoF$_3$ crystal has all conditions to be an AM.

  To determine if the ground state of the system is spin-polarized, we analyzed the energy difference between spin-polarized and non-spin-polarized calculations. The results, shown in Fig. S2, indicate that the material exhibits magnetic order, with the energy difference significantly increasing with the Hubbard term. The involvement of the cobalt atom's d orbitals near the Fermi level underscores the significance of the Hubbard term and correlation effects. The band gap size of the structure is also directly influenced by the parameter U, as illustrated in Fig. S3. An increase in U results in a larger band gap. Experimental studies have demonstrated that the CoF$_3$ structure is an antiferromagnet with a Neel arrangement at low temperatures. To investigate the magnetic ground state of the system, we considered both ferromagnetic and antiferromagnetic Neel configurations. In the antiferromagnetic configuration, the spin alignment of each cobalt plane is along the Z direction, with alternating planes rotated. The spin difference, $\Delta E=E_{AFM}-E_{FM}$, is depicted in Fig. S4. As shown, for U<3, the two spin states are degenerate. Increasing U leads to the formation of the antiferromagnetic ground state. This finding aligns with previous calculations that suggested U=5.3 eV for cobalt.

\begin{figure}
	\includegraphics[width=0.45\textwidth]{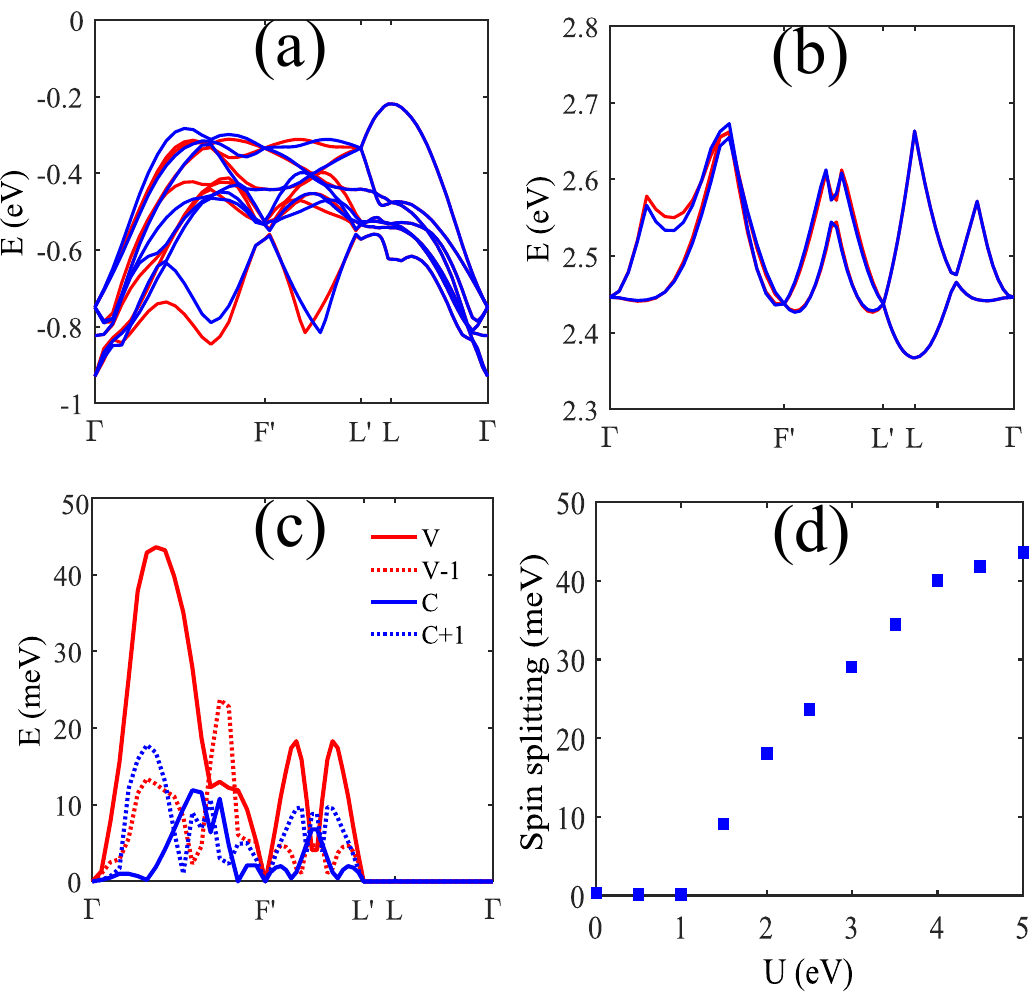}
	\caption{\label{Fig3}Spin-polarized band structure of CoF$_3$ for (a) valence and (b) conduction bands. Spin-up (-down) is shown y blue (red) lines. (c) Spin splitting of the valence band (V), the band below the valence band (V-1), the conduction band (C), and the band above the conduction band (C+1). (d) Variation of the maximum spin splitting as a function of the Hubbard term.}
\end{figure}

The spin band structure of CoF$_3$ with U=5 eV for the valence and conduction bands is shown in Fig. \ref{Fig3}a and b, respectively. Despite the antiferromagnetic arrangement of the structure, which results in zero net magnetization of the unit cell, we observe a breaking of time-reversal symmetry and the loss of spin Kramers degeneracy. This spin splitting is evident in both the valence and conduction bands and can reach up to approximately 45 meV. The spin splitting in the valence band is slightly greater than in the conduction band. To highlight the antiferromagnetic (AFM) nature of our sample, we have separately plotted the valence and conduction band structures. Results for other U values are shown in Fig. S5.  For U > 2 eV, AFM behavior appears in the sample, and the breaking of Kramers degeneracy increases directly with U. In Fig. \ref{Fig3}c, we have plotted the difference in the band structure of spin-up and spin-down in the first Brillouin zone. At the $\Gamma$ point, Kramers degeneracy is preserved. Additionally, we observe spin degeneracy along the $\Gamma$-L-F path. In fact, the AM characteristics are not observed in the $k_z=0$ plane. The maximum spin splitting occurs along the $\Gamma$-F’ path. It is also evident that the spin splitting depends on the atomic orbitals forming the band. Among the four considered bands, which include VB-1, VB, CB, and CB+1, the highest spin splitting is observed in the VB band. Figure \ref{Fig3}d demonstrates the dependence of the maximum spin splitting on the Hubbard term. As shown, the spin splitting significantly increases with the Hubbard term.

The observed spin splitting in the structure is due to the electric crystal potential. The magnitude and momentum dependence of the unusual spin splitting in this material are determined by the electric crystal potential of its non-magnetic phase. For the electric crystal potential to lead to spin splitting and the emergence of the AM phase, the magnetic crystal must be anisotropic. This anisotropy allows the $[C_2||\mathbf{G-H}]$ symmetries to separate opposite spins at the same energy in momentum space. Additionally, the symmetries that exchange identical spins in the lattice, $[E||\mathbf{H}]$, must be sufficiently low to create appropriate anisotropy in momentum space. The crystallographic symmetry group ($\overline{3}m$) and the non-trivial spin group ($\overline{3}^2m$) provide these conditions for the CoF$_3$ crystal to exhibit significant spin splitting in the $k_z\neq 0$ plane. Thus, we can define a model Hamiltonian considering the symmetries of the material as follows:
\begin{equation}
H=\Delta \sigma_z k_zk_x(k_x^2-(\sqrt{3}k_y)^2),
\end{equation}
where $\Delta$ is spin splitting, and $\sigma_z$ denotes the pauli matrix. The above Hamiltonian indicates that the spin splitting is maximized in the ($k_z=0.5$) plane, which is consistent with the DFT results. However, the spin splitting in this structure is weaker compared to MnTe \cite{osumi2024observation, lee2024broken} and RuO$_2$ \cite{vsmejkal2023chiral, fedchenko2024observation} structures, due to the weaker electric crystal potential of this material. In fact, the other two materials, with their highly asymmetric spin density, have more favorable conditions for the removal of Kramers spin degeneracy.

\begin{figure}
	\includegraphics[width=0.45\textwidth]{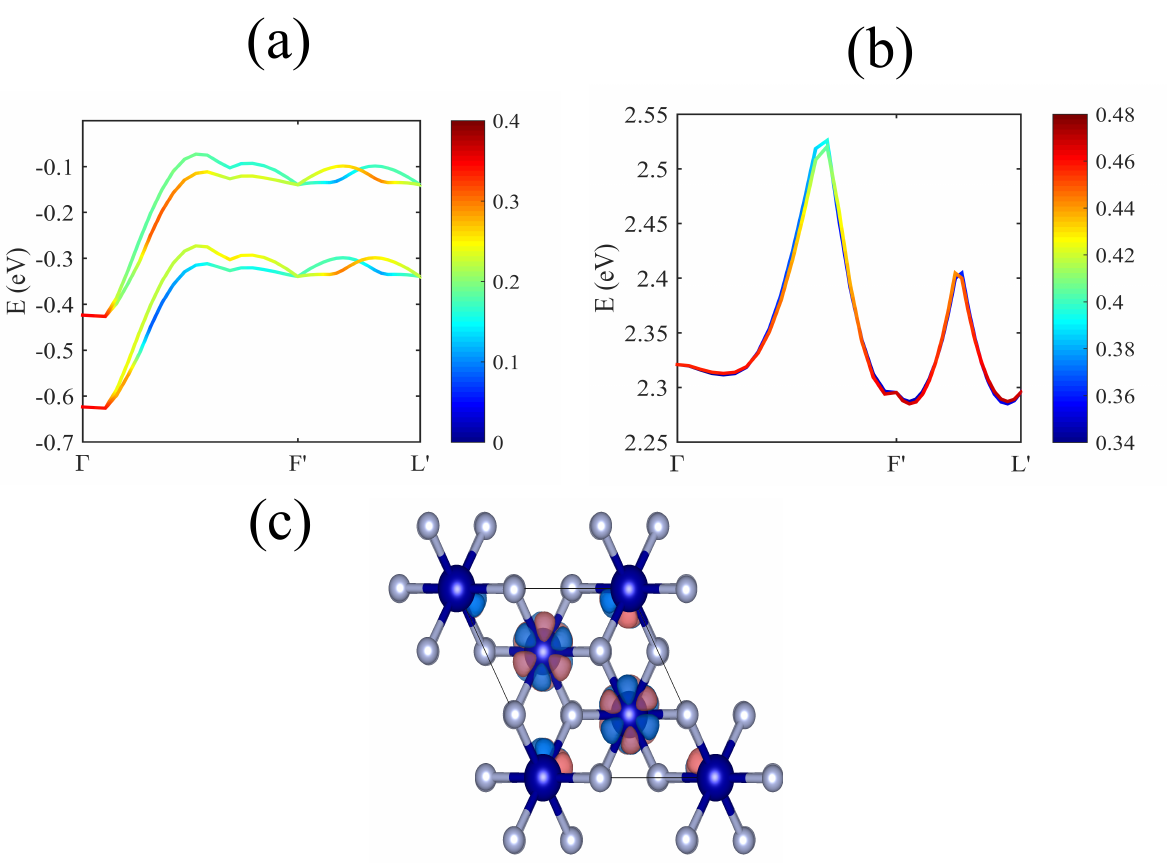}
	\caption{\label{Fig4} (a) and (b) The valence and conduction bands projected onto two cobalt atoms corresponding to two different planes along the z-axis, respectively. To better illustrate the band structure, the projected band structure of one atom has been shifted by 0.2 eV. (c) Spin-momentum locking, indicative of the g-wave AM structure.}
\end{figure}

To achieve a more comprehensive understanding of the influence of symmetries on the emergence of AM, we have illustrated the valence and conduction bands  projected on two cobalt atoms situated on different $k_z$ planes and adjacent to each other in Fig.\ref{Fig4}. The results indicate that along the $\Gamma$-F’ path in the valence band, the energy levels of the spin-up (-down) states in the two atoms exhibit unequal weights so that change of the plane results in the dominance of one of the spin states. This finding contrasts with the results observed in an antiferromagnet, where equal weight contributions are present for the cobalt atoms. Along the $F’$-$L’$ path, the dominance of spin weight undergoes a reversal in spin direction. In the conduction band, the disparity in spin weight between the bands is significantly reduced. Notably, the AM behavior in the valence band is more pronounced, attributed to the strong hybridization between the p orbitals of the chlorine atom and the d orbitals of the cobalt atom. This interaction markedly enhances the anisotropy of the spin density of the cobalt atoms on different planes with varying spin orientations, leading to the breaking of time-reversal symmetry. Additionally, spin-momentum locking is depicted in Fig. \ref{Fig4}c. The figure indicates the presence of a bulk g-wave. Furthermore, a 60$^0$ spin rotation is necessary to exchange cobalt atoms with differing spins, which is a fundamental aspect of altermagnetism.

    This study underscores the significance of altermagnetism as a novel magnetic phase with unique properties that bridge the gap between ferromagnetism and antiferromagnetism. Through density functional theory (DFT) and spin group theory, we have demonstrated that cobalt trifluoride (CoF$_3$) exhibits characteristics of altermagnetism, including zero net magnetization and significant spin splitting without spin-orbit coupling. The strong hybridization between the p orbitals of chlorine and the d orbitals of cobalt enhances the anisotropy of spin density, further contributing to the breaking of Kramers degeneracy. The findings highlight the role of crystallographic symmetries and electric crystal potentials in stabilizing this phase, leading to robust time-reversal symmetry breaking and spin-momentum locking.
The unique electronic properties of altermagnets, such as those observed in CoF$_3$, present promising opportunities for advancements in spintronic devices, high-density data storage, and quantum computing.

\begin{acknowledgments}
I am grateful to the Research Council of the University of
Guilan for partial support of this research.
\end{acknowledgments}

\section*{Conflict of Interest}
The authors have no conflicts to disclose

\section*{Author Contributions}
M. B. Tagani performed the simulations, analyzed the data, wrote the initial draft and finalized the article. 

\section*{Data Availability Statement}

The data that support the findings of
this study are available from the
corresponding author upon reasonable
request.

\bibliography{Ref}

\end{document}